\documentclass[a4paper,11pt]{article}
\usepackage{jinstpub} 
\RequirePackage{graphicx}
\RequirePackage{siunitx}
\usepackage{comment}
\usepackage{amsmath} 
\usepackage{array}
\usepackage{upgreek,xspace}
\def\textalpha{\ensuremath\upalpha}
\def\textmuon{\ensuremath\upmu}
\newcommand{\fprompt}{F$_{\mathrm{prompt}}$}
\newcommand{\fpromptm}{\mathrm{F}_{\mathrm{prompt}}}
\newcommand{\nuc}[2]{\ensuremath{^{#2}\mbox{#1}}}

\usepackage{subcaption}
\title{Demonstration of the light collection stability of a PEN-based wavelength shifting reflector in a tonne scale liquid argon detector
}

\author[f]{V.~Gupta,}
\author[i]{G.\,R.~Araujo,}
\author[i]{M.~Babicz,}
\author[i]{L.~Baudis,}
\author[i]{P.-J.~Chiu,}
\author[a]{S.~Choudhary,}
\author[g]{M.~Goldbrunner,}
\author[d]{A.~Hamer,\footnote{now at STFC Boulby Underground Laboratory, Boulby, TS13 4UZ, UK}}
\author[a]{M.~Ku\'zniak,}
\author[a]{M.~Ku\'zwa,}
\author[g]{A.~Leonhardt,}
\author[b]{E.~Montagna,}
\author[a]{G.~Nieradka,}
\author[d]{H.\,B.~Parkinson,}
\author[c,h]{F.~Pietropaolo,}
\author[f]{T.\,R.~Pollmann,}
\author[c]{F.~Resnati,}
\author[g]{S.~Sch\"onert,}
\author[d]{A.\,M.~Szelc,}
\author[e]{K.~Thieme}
\author[a]{and M.~Walczak\footnote{now at Gran Sasso Science Institute, L’Aquila 67100, Italy and  INFN Laboratori Nazionali del Gran Sasso, Assergi (AQ) 67100, Italy}}

\affiliation[a]{AstroCeNT, Nicolaus Copernicus Astronomical Center of the Polish Academy of Sciences, Rektorska 4, 00-614 Warsaw, Poland}
\affiliation[b]{INFN, Sezione di Bologna, 40127 BO, Italy}
\affiliation[c]{CERN, The European Organization for Nuclear Research, 1211 Meyrin, Switzerland}
\affiliation[d]{University of Edinburgh, Edinburgh EH9 3FD, United Kingdom}
\affiliation[e]{Department of Physics and Astronomy, University of Hawai’i, Honolulu, HI 96822, USA}
\affiliation[f]{Nikhef and the University of Amsterdam, Science Park, 1098XG Amsterdam, The Netherlands}
\affiliation[g]{TUM School of Natural Sciences, Department of Physics, Technical University of Munich, James-Franck-Str. 1, Garching, Germany}
\affiliation[h]{INFN, Sezione di Padova, via Marzolo 8, 35131, Italy}
\affiliation[i]{Physik-Institut, University of Zurich, Zurich, Switzerland}

\abstract{Liquid argon detectors rely on wavelength shifters for efficient detection of scintillation light. The current standard is tetraphenyl butadiene (TPB), but it is challenging to instrument on a large scale. Poly(ethylene 2,6-naphthalate) (PEN), a polyester easily manufactured as thin sheets, could simplify the coverage of large surfaces with wavelength shifters. Previous measurements have shown that commercial grades of PEN have approximately~50\% light conversion efficiency relative to TPB. Encouraged by these results, we conducted a large-scale measurement using~\SI{4}{\square\metre} combined PEN and specular reflector foils in a two-tonne liquid argon dewar to assess its stability over approximately two weeks. This test is crucial for validating PEN as a viable substitute for TPB. The setup used for the measurement of the stability of PEN as a wavelength shifter is described, together with the first results, showing no evidence of performance deterioration over a period of 12 days.}

\keywords{wavelength shifters, polyethylene naphthalate, PEN, liquid argon detectors}

\begin{document}
\maketitle

\section{Introduction} \label{intro}

Detectors using liquid argon (LAr) as active target material are common in dark matter and neutrino physics experiments~\cite{boniventoScienceTechnologyLiquid2024}. These detectors rely on wavelength-shifting (WLS) materials, in particular tetraphenyl butadiene (TPB), to convert argon scintillation light produced at \SI{128}{\nano\metre} to visible wavelengths, enabling efficient detection~\cite{instruments5010004,DarkSide,detectorpaper,protodune,microboone,ArDM,coherent2018,LAr_inst_proceeding,pcdr_LEGEND:2021bnm,DS20K}. To obtain a uniform and efficient WLS layer, TPB must be applied on a substrate by vacuum deposition~\cite{Canci_2024,Broerman:2017hf}, which makes it difficult to use in the upcoming generation of multi-tonne-scale detectors with surface areas of many hundreds of \SI{}{\square\metre}, much larger than a typical vacuum evaporator~\cite{pcdr_LEGEND:2021bnm,DS20K}. 
Poly(ethylene 2,6-naphthalate) (PEN) is currently considered a replacement for TPB~\cite{nakamuraEvidenceDeepbluePhoton2011, kuzniakPolyethyleneNaphthalateFilm2019,boulayDirectComparisonPEN2021,araujoWavelengthshiftingReflectorsCharacterization2022} and for use as a self-vetoing structural material~\cite{efremenkoUsePolyEthylene2020,hackettLightResponsePoly2022}. PEN has been shown to have 30\% to 75\% of the efficiency of TPB for shifting LAr scintillation photons to visible wavelengths~\cite{araujoWavelengthshiftingReflectorsCharacterization2022,kuzniakPolyethyleneNaphthalateFilm2019,boulayDirectComparisonPEN2021,abrahamWavelengthShiftingPerformancePolyethylene2021,abudScintillationLightDetection2022}, with the measured relative light yield (LY) depending strongly not just on the intrinsic quantum efficiency of the WLS grade and batch but also on the configuration of the setup, such as geometry and photosensor coverage, as well as the optical properties of the structural carrier, reflector, and WLS layers.

So far, only small samples of less than \SI{0.04}{\square\metre} of PEN have been studied in LAr test setups~\cite{araujoWavelengthshiftingReflectorsCharacterization2022,boulayDirectComparisonPEN2021,abrahamWavelengthShiftingPerformancePolyethylene2021}: only one measurement was carried out for more than a few days~\cite{abudScintillationLightDetection2022}, with no explicit stability results reported. We report here on the uniformity and stability of a \SI{4}{\square\metre} PEN-based wavelength shifting reflector (WLSR) setup in a LAr detector over a period of 12~days, by measuring the LY using \nuc{Am}{241} \textalpha's and cosmic ray (CR) secondary \textmuon's. 
This is an essential qualification step for using PEN as the wavelength shifter in the next generation of LAr experiments.

\section{Experimental setup} \label{sec:setup}
\begin{figure}
         \centering
         \includegraphics[width=0.45\columnwidth]{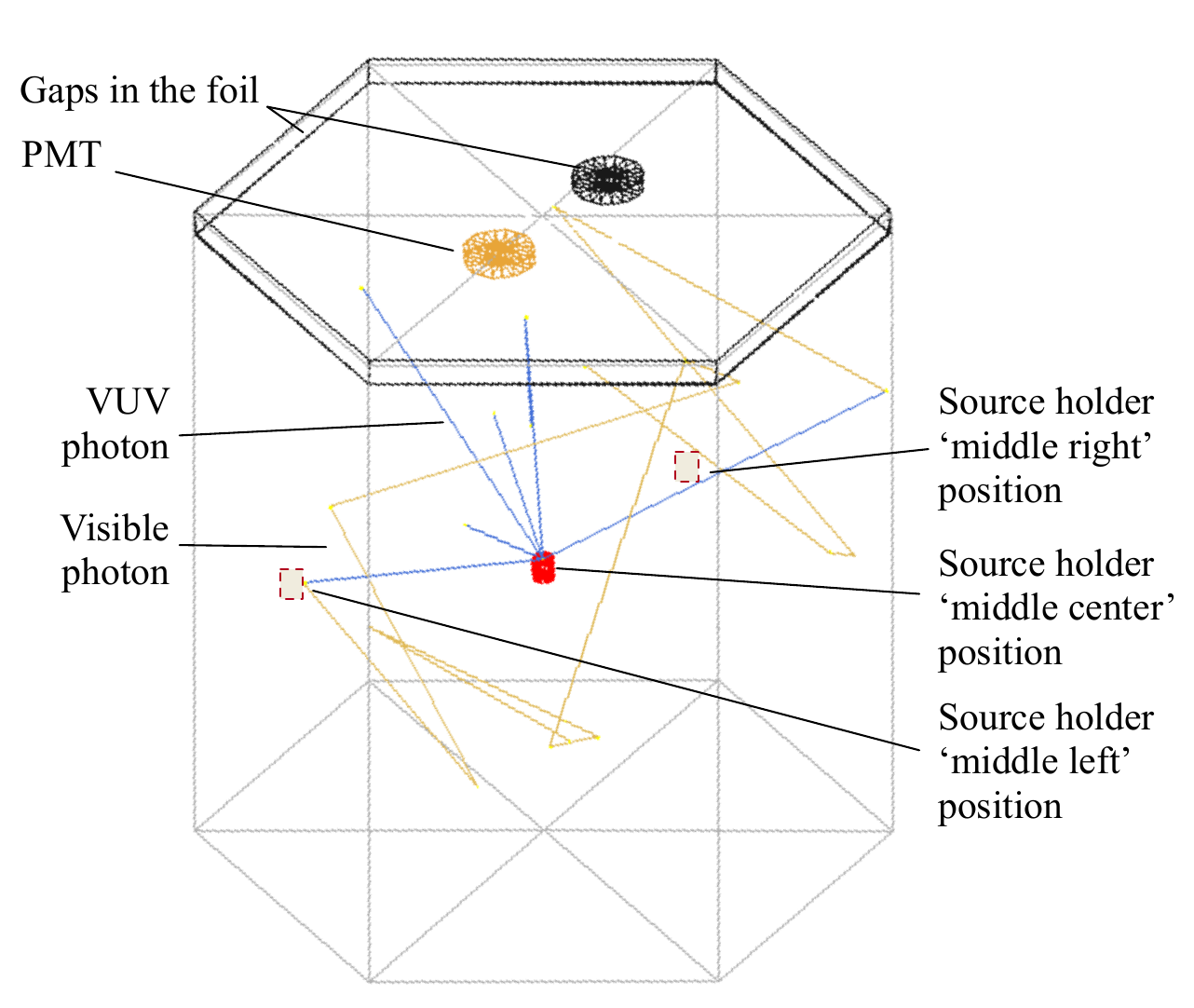}~\includegraphics[width=0.28\textwidth]{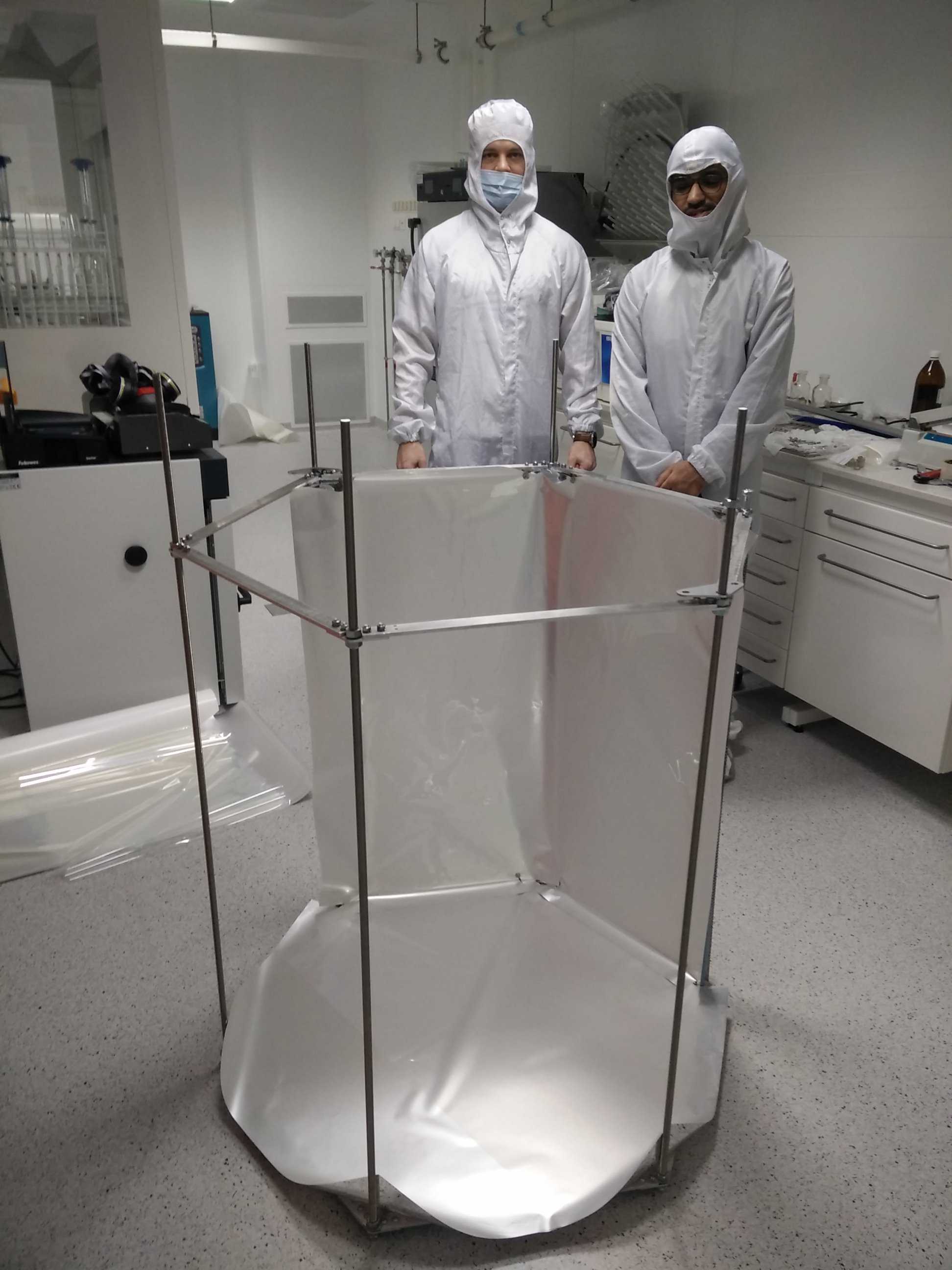}~\includegraphics[width=0.28\textwidth]{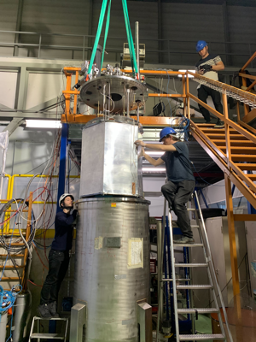}
        \caption{(left) Experimental setup as implemented in GEANT4. Scintillation light, produced when \textalpha\ particles deposit energy in LAr, is wavelength-shifted to visible light at the PEN walls. Visible photons reflect on the surfaces until they are either absorbed or detected by the PMT. The \textalpha\ source is moved radially and vertically to different positions for characterizing foil uniformity. 
        (middle) Picture of the hexagonal cage partially lined with PEN and reflector foils. (right) Installation of the cage in the 2-tonne LAr dewar at CERN.}
        \label{fig:wlssetup}
\end{figure}

\subsection{PEN WLSR setup}
A conceptual design of the experimental setup is shown in figure~\ref{fig:wlssetup}~(left). We use commercially available \SI{25}{\micro\metre} thick Teonex Q53 PEN based on measurements reported in refs.~\cite{boulayDirectComparisonPEN2021,araujoWavelengthshiftingReflectorsCharacterization2022} loosely backed by a 3M DF2000MA specular reflector film.\footnote{The adhesive backing layer was wetted with toluene and manually removed. The remaining film is similar to 3M Enhanced Specular Reflector (ESR).} The WLSR foils are assembled on the inner surface of a hexagonal cage with an outer diameter of \SI{99.5}{\centi\metre} and a height of \SI{101.5}{\centi\metre}, and installed inside a two-tonne LAr dewar as shown in figure~\ref{fig:wlssetup}~(middle, right), located in the CERN Building 182. 
An \nuc{Am}{241} \textalpha\ source is placed on the tip of an L-shaped lever connected to a linear and rotary motion feedthrough (also lined with the reflector) such that it can be moved vertically and laterally within the cage. The \textalpha\ emission rate from the surface of the source was assayed to be \SI{17.8 \pm 1.4}{\kilo\hertz}. The source emits \textalpha\ particles with a peak energy of \SI{4.8}{\mega\eV},\footnote{The peak energy is less than the nominal \SI{5.5}{\mega\eV} due to a \SI{1.8}{\micro\metre} palladium protective film covering the radioactive material, in which the \textalpha\ particles lose some of their energy. The \SI{60}{\kilo\eV} gamma emitted from \nuc{Am}{241} has a negligible contribution to the overall energy deposited in LAr.} which have a range of approximately \SI{43}{\micro\metre} in LAr. For measuring the LY uniformity, the \textalpha\ source is placed at a height of approximately \SI{80}{\centi\meter} ("top"), \SI{40}{\centi\metre} ("middle"), and \SI{0}{\centi\metre} ("bottom") with respect to the bottom surface of the cage. Additionally, the source is also rotated horizontally to \SI{0}{\degree}~("center"), \SI{+50}{\degree}~("right"), and \SI{-50}{\degree}~("left"). The photomultiplier tube (PMT) is located at an angle of \SI{-20}{\degree} with respect to the "center" position. A white LED is also placed on the \textalpha\ source holder.

As the setup is on ground level, CR \textmuon's also deposit energy in the LAr volume. The \textmuon\ spectrum peaks in the energy range of $\SIrange{0.1}{10}{\giga\eV}$ where the \textmuon's are minimum ionizing particles (MIP). With CR angles of incidence peaking toward the vertical direction, most \textmuon's travel the length of the detector, or approximately \SI{1}{\metre}, depositing approximately \SI{300}{\mega\eV} along their path.

LAr is supplied from an external supply tank with a $\mathcal{O}(\SI{1}{ppm})$ impurity level. The dewar is operated at $\SIrange{40}{100}{mbarg}$ with a gas ullage of $\SIrange{3}{15}{\%}$. A purifier from ICARUS~\cite{Cennini:1993abz,Vignoli:2015jxa} consisting of a $\SI{2}{\liter}$ molecular sieve (Hydrosorb) and a $\SI{4}{\liter}$ chromium oxide cartridge (Oxisorb\textsuperscript{\textregistered}), for water and oxygen adsorption, is located on the top flange and open to the gas ullage. Regular refills with the filter in-line ensure a minimum LAr level is maintained, monitored with a capacitive level meter, such that the PMT and its electronics base are fully submersed in the liquid.

\subsection{Light detection and data acquisition}
\begin{figure}
     \centering
     \includegraphics[width=0.32\columnwidth]{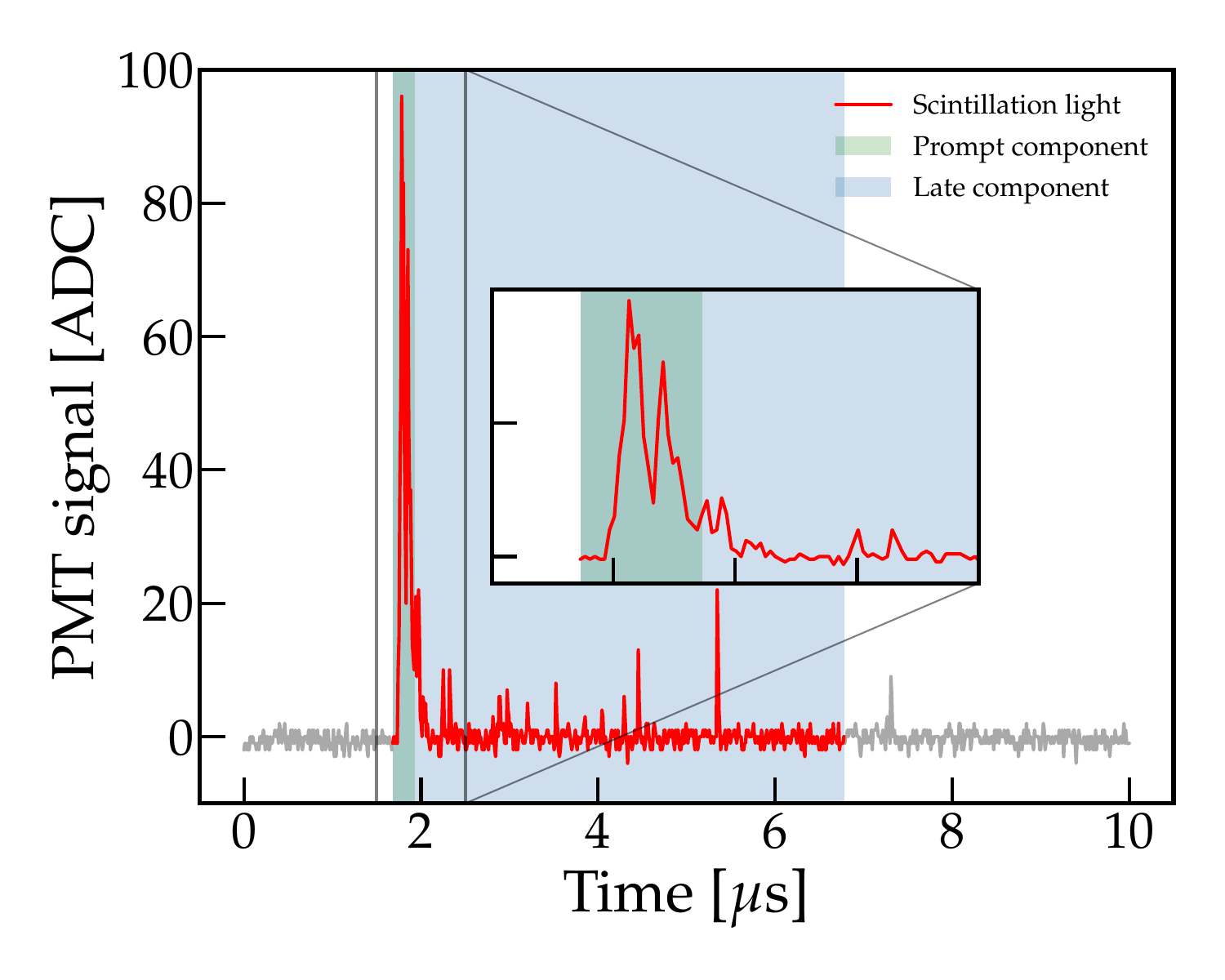}~\includegraphics[width=0.32\columnwidth]{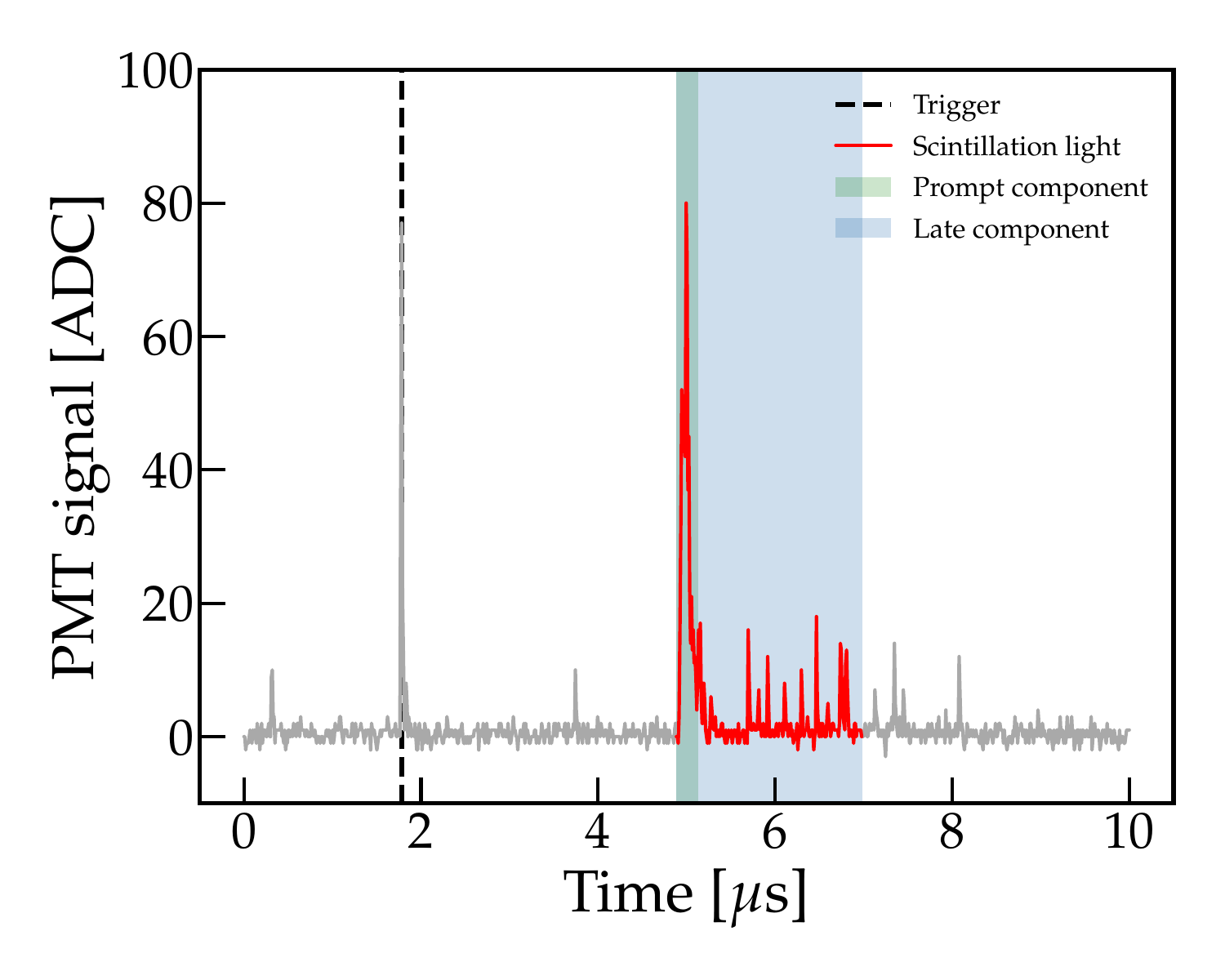}~\includegraphics[width=0.32\columnwidth]{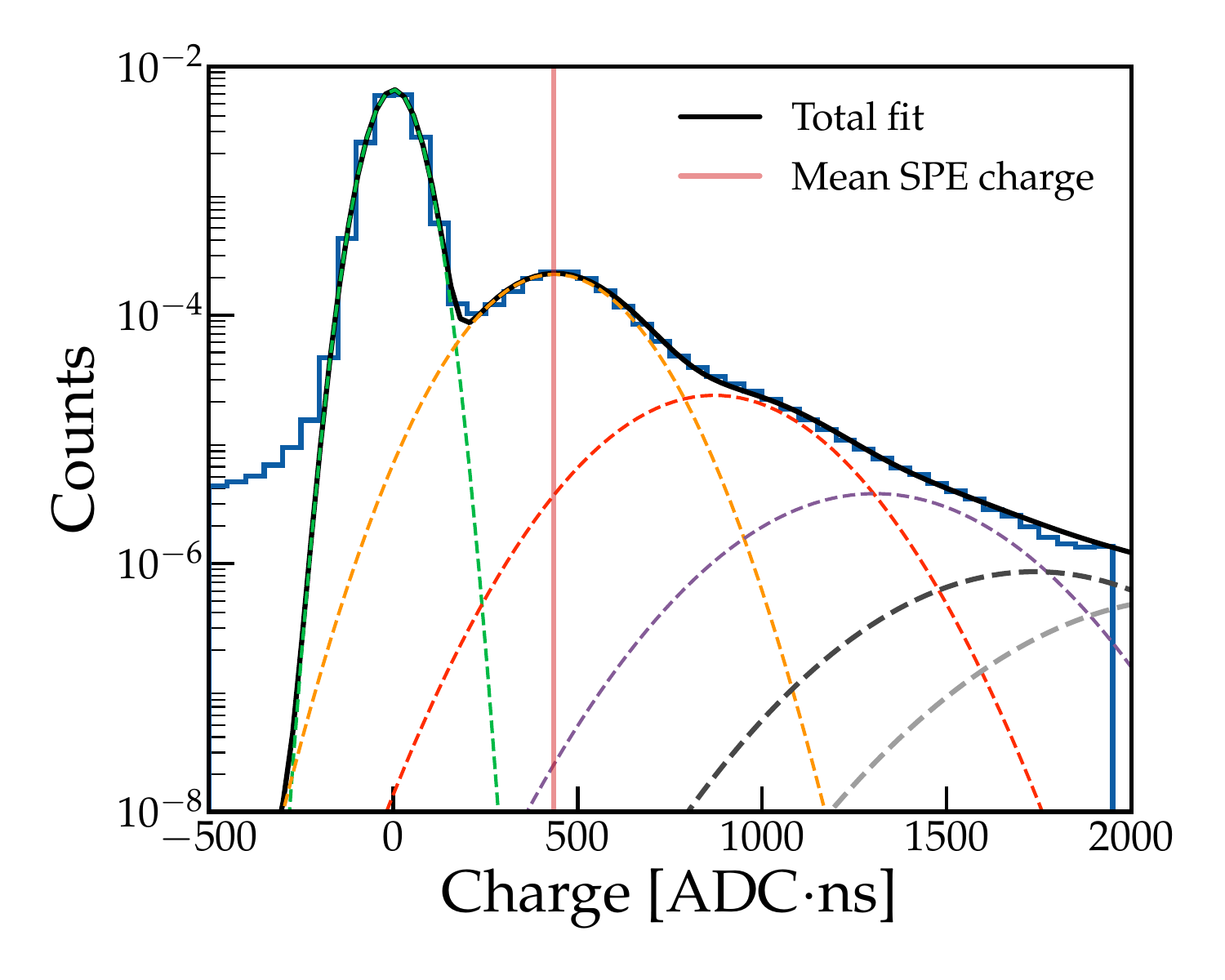}
    
    \caption{Traces of \textalpha-events, triggering the DAQ (left) or in the post trigger window (middle). The energy of the \textalpha\ event is estimated by the area of pulses in the blue shaded time window. The green-shaded region indicates the prompt time window. (right) SPE spectrum with model fit.\label{fig:traces}}
\end{figure}

The wavelength-shifted light is detected by an 8" VUV-blind Hamamatsu R5912-MOD PMT, repurposed from the ICARUS experiment~\cite{Babicz_2018} and installed at the top of the cage, looking into the cage volume through a circular hole of 4" diameter in the WLSR foil.
At the PEN peak emission wavelength of \SI{430}{\nano\metre}~\cite{maryUnderstandingOpticalEmissions1997}, the PMT has a quantum efficiency of 16--18\%~\cite{Babicz_2018,zhaoMeasurementRelativeQuantum2021}. The PMT signals are recorded with a 14-bit CAEN V1724 digitizer with a $\SI{100}{MHz}$ bandwidth, set to self-trigger. The trigger threshold was set to pass the \nuc{Am}{241} \textalpha\ events while rejecting the lower-energy background events from \nuc{Ar}{39} beta decays.
When a trigger happens, the PMT voltage is converted to ADC units with a sampling rate of \SI{100}{MS/\second} and saved to disk with a pre-trigger window of \SI{2}{\micro\second} and a post-trigger window of \SI{8}{\micro\second}. An example trigger of an \textalpha\ event is shown in figure~\ref{fig:traces}~(left). The detector was filled with LAr on January 18, 2023, and data was taken between January 24, 2023 and February 10, 2023. An oscilloscope was used for the first four days of data taking due to malfunction of the original data acquisition system, after which a new DAQ was installed and normal data taking resumed.

\section{Data analysis and event discrimination} \label{sec:analysis}
The peak-finding algorithm identifies signals with height exceeding 50\% of a single photoelectron (SPE) peak. The SPE spectrum is found by integrating low-amplitude peaks in the data, while the pedestal is constructed by integrating over regions with no peaks. The mean SPE charge is estimated by fitting a sum of Gaussian distributions to the SPE spectrum, as shown in figure~\ref{fig:traces}~(right). The mean SPE charge drifted downwards by 1\% per day.

\begin{figure}[h]
     \centering
    \includegraphics[width=0.45\columnwidth]{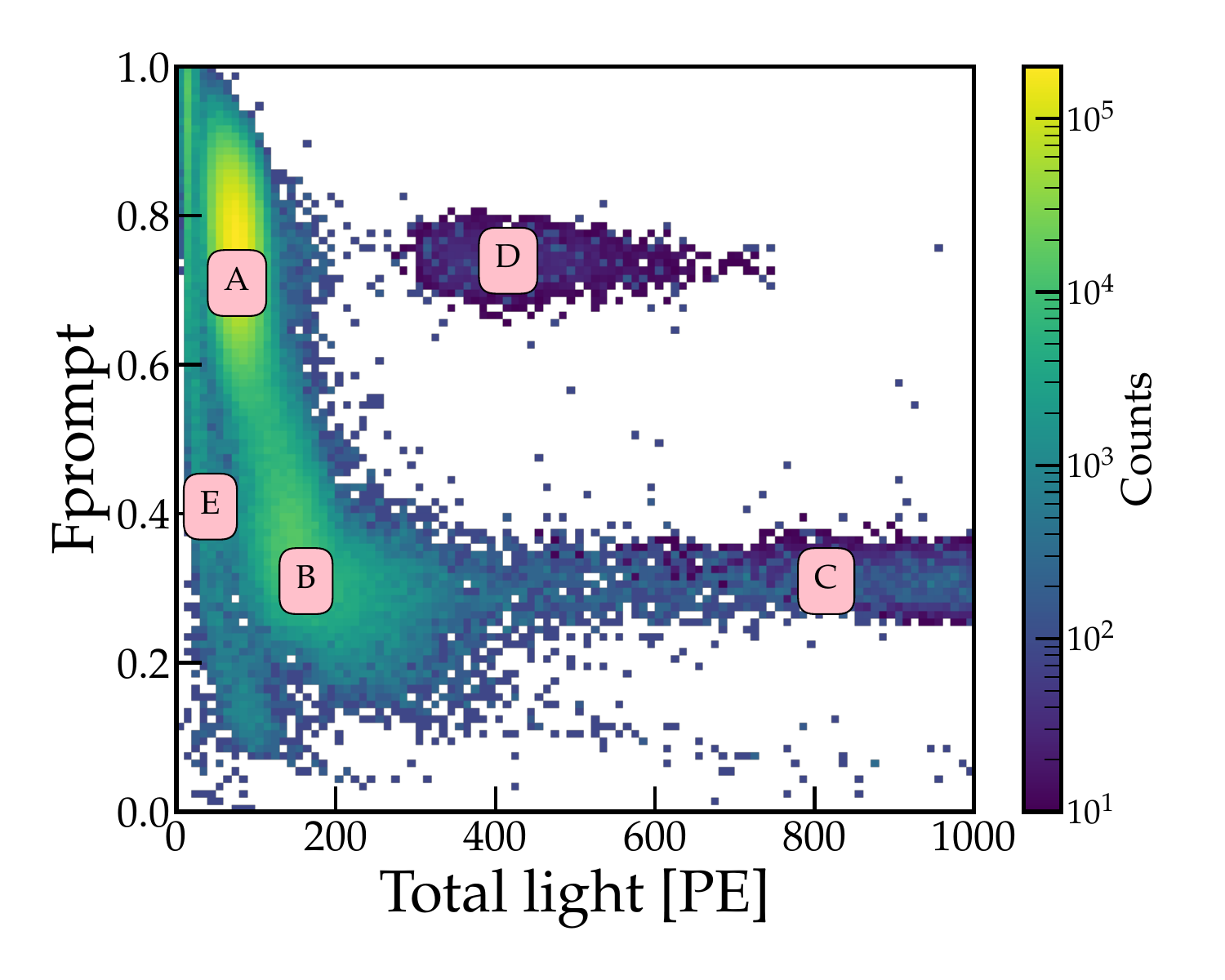}~\includegraphics[width=0.45\columnwidth]{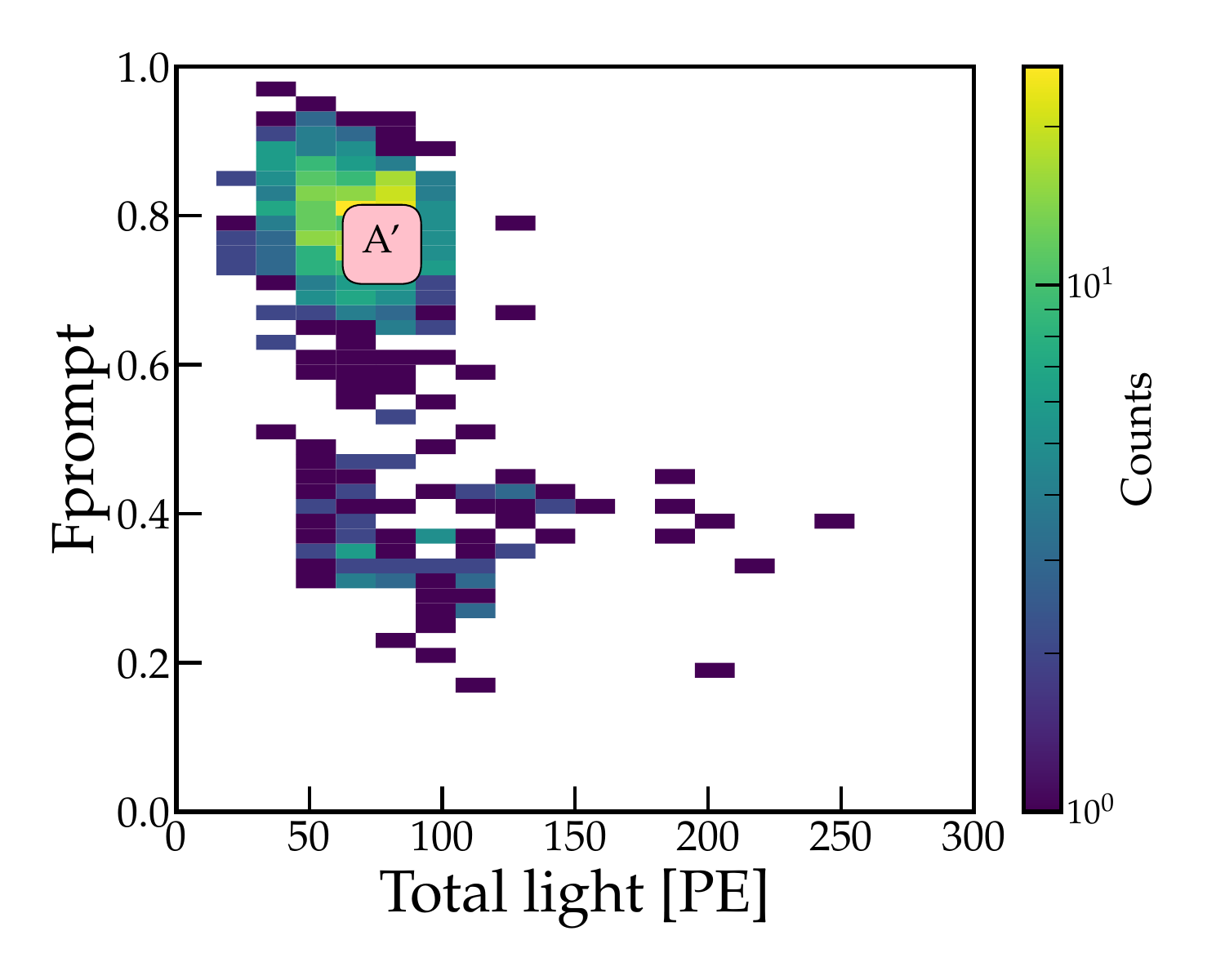}
    \caption{Events, occurring at the trigger time (left) or in the post-trigger window (right), recorded over the span of 1 hour in Fprompt vs.~TotalPE space. Different populations are identified:  \textbf{A} - \nuc{Am}{241} \textalpha's. \textbf{B} - \nuc{Am}{241} \textalpha's with pileup. \textbf{C} - CR \textmuon's. \textbf{D} - Unkown population but most likely candidate is protons from inelastic interaction of CR \textmuon's and neutrons. \textbf{E} - Short bursts of light inconsistent with scintillation; several mechanisms create such $\mathcal{O}$(10~PE) pulses, such as Cherenkov light produced in the PMT cathode glass, PMT afterpulsing, and ionizing particles (from CR's or \nuc{Ar}{39} beta decays) interacting in the PEN foil. \fprompt\ for these events is close to 1, and the band stretching toward lower values is due to pile-up. \textbf{A$^{\prime}$} - \nuc{Am}{241} \textalpha's in post-trigger window of \textbf{E}. \label{fig:one_subrun}}
    
\end{figure}

Scintillation light in LAr has a double-exponential time structure due to the excimers in the singlet and triplet states decaying with an approximately \SI{6}{\nano\second} and \SI{1400}{\nano\second} lifetime, respectively~\cite{Carvalho:1979tm,Kubota:1978kh,Morikawa:1989gv,Hofmann:2013hf}. The ratio of the number of singlet and triplet excimers created can be used for particle identification~\cite{boniventoScienceTechnologyLiquid2024}. This ratio is estimated using the \fprompt\ parameter, defined as the ratio of light detected in a prompt time window of [$T_0$-\SI{100}{\nano\second}, $T_0$+\SI{150}{\nano\second}] and a total time window of [$T_0$-\SI{100}{\nano\second}, $T_0$+\SI{5000}{\nano\second}] (TotalPE), where $T_0$ is the trigger time. Figure~\ref{fig:one_subrun} shows the events recorded in one hour in \fprompt\ vs. TotalPE space and several event populations.

After preliminary analysis, it was found that the trigger threshold in many runs was set at a level where part of the \textalpha\ population (A) was cut off. To obtain an unbiased sample of the \nuc{Am}{241} \textalpha's, the events in population (E) were used as a random trigger sample. As can be seen in figure~\ref{fig:traces}~(middle), these events have a short flash of light at the trigger time, and possibly another event later in the trace. We discard the trigger time region and then apply a software trigger that sets the trigger time to the largest peak in the rest of the trace. To minimize the effect of pile-up, we remove events that have another peak whose size is more than 10\% of the prompt peak. Figure~\ref{fig:one_subrun}~(right) shows the distribution of  events selected this way in \fprompt~and TotalPE space.  

\begin{figure}
     \centering
    \includegraphics[width=0.3\columnwidth]{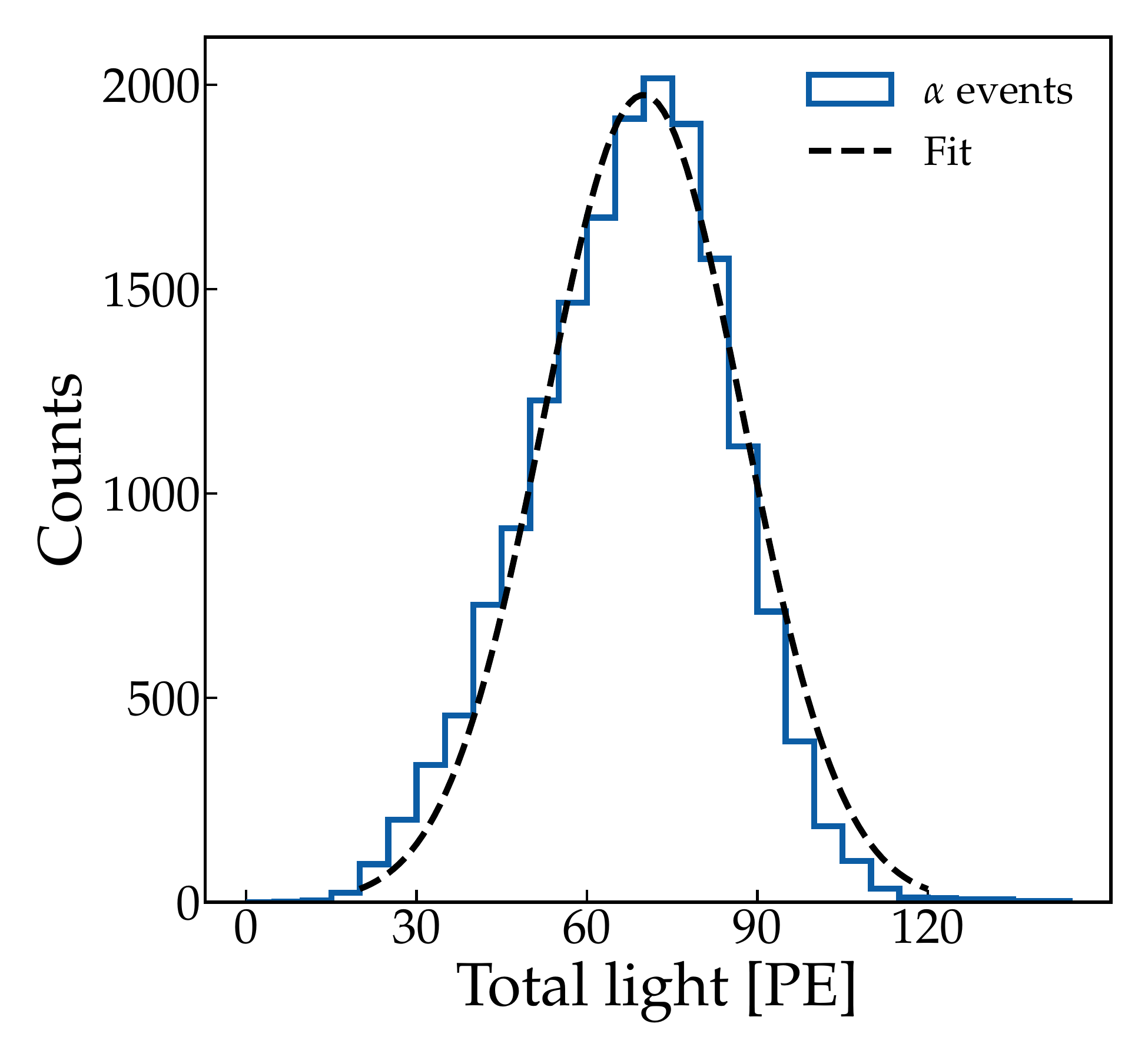}~\includegraphics[width=0.3\columnwidth]{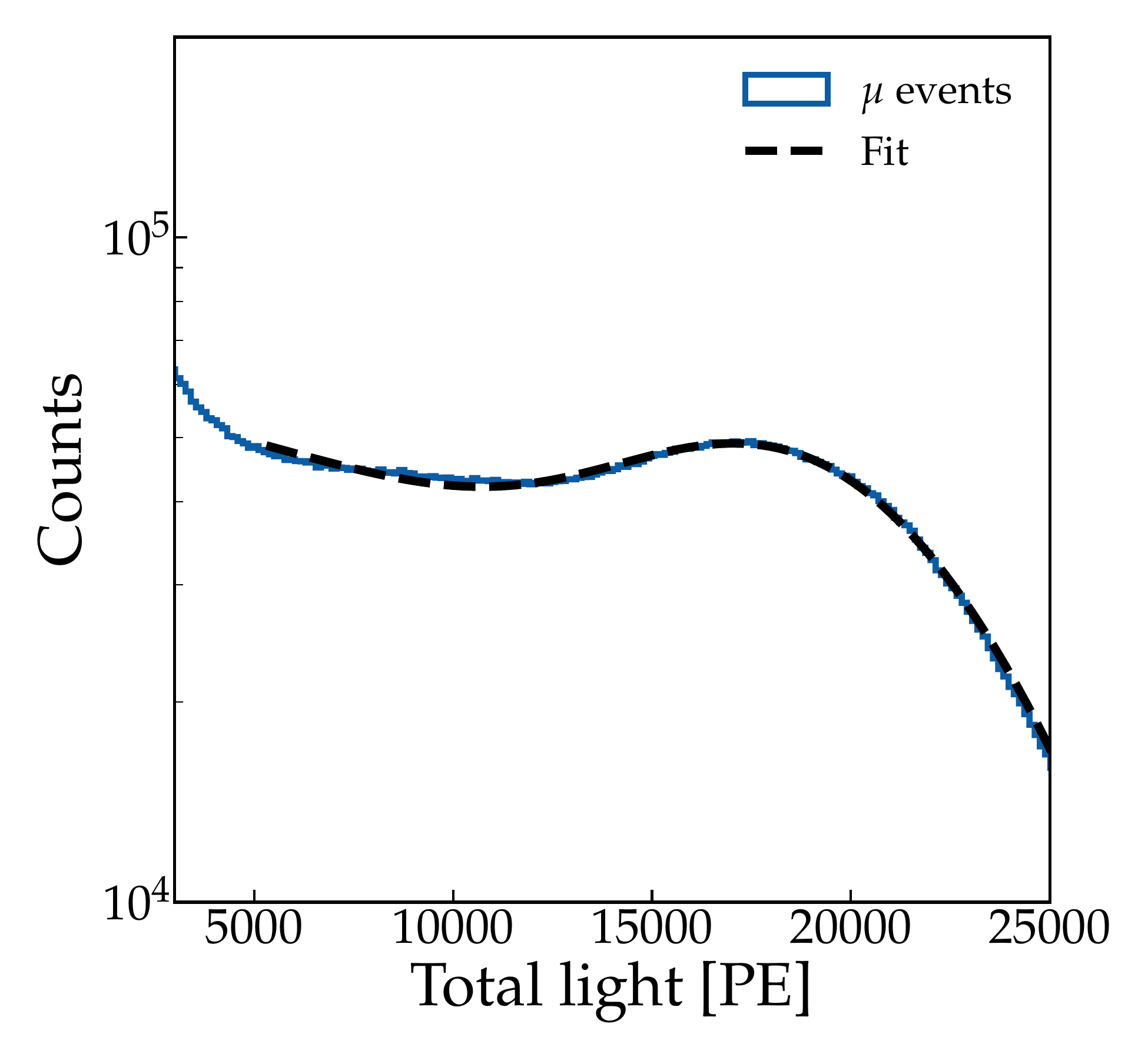}~\includegraphics[width=0.3\columnwidth]{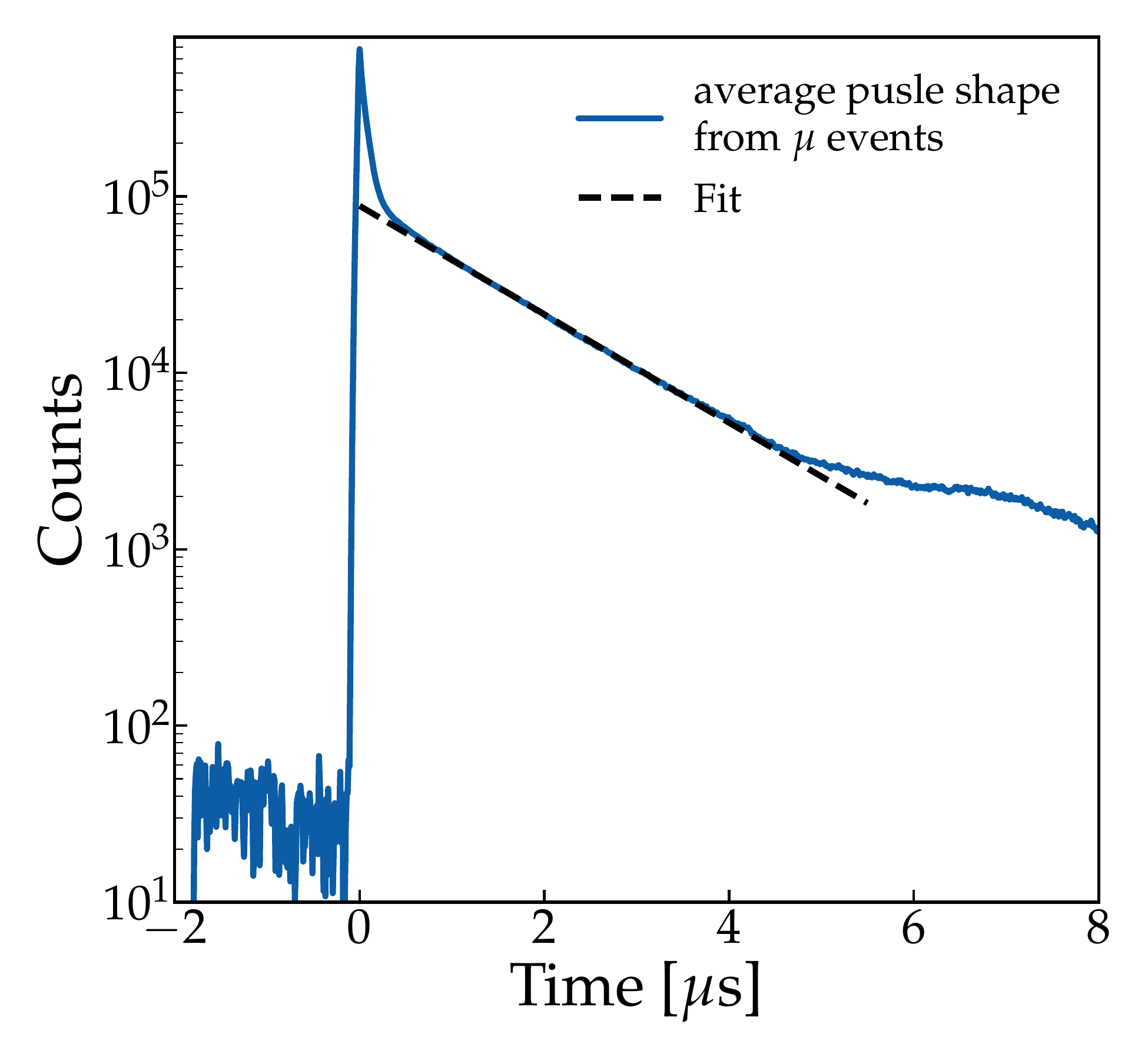}
    \caption{Example energy spectra recorded for (left)~\nuc{Am}{241} \textalpha's and (middle)~CR secondary \textmuon's. (right) LAr pulse shape from \textmuon\ events. \label{fig:alpha_fit}}
\end{figure}

We track the LY using \nuc{Am}{241} \textalpha's (A/A$^{\prime}$) and CR secondary \textmuon's (C). Example fits of the \textalpha\ spectrum with a Gaussian function and of the \textmuon\ spectrum with a Gaussian function on top of a linearly falling background are shown in figure~\ref{fig:alpha_fit}~(left, middle). We measured the LAr triplet lifetime, $\tau_t$, daily, by summing the traces of \textmuon\ events to create an average pulseshape. The triplet lifetime is estimated by fitting an exponential distribution 
$
    I(t) = \frac{I_0}{\tau_t} e^{-t/\tau_t} + c
$
to the pulseshape in the time range of $\SIrange{1}{4}{\micro\second}$ after the trigger, as shown in figure~\ref{fig:alpha_fit}~(right).  

The LY is calculated using the peak positions $m$ from the spectrum fits, measured in PE, divided by the electron-equivalent visible energy:
\begin{align} \label{eq:ly}
    \text{LY}_\alpha = \frac{m_\alpha}{E_\alpha \cdot q_\alpha \cdot \eta_{\text{source}}}  & \ ,\ \ \  
        & \text{LY}_\mu = \frac{m_\mu}{E_\mu} ,
\end{align}
where $E_\alpha$\,/\,$E_\mu$ is the energy of the \textalpha\,/\,\textmuon\, particle, 
$q_\alpha=0.72(0.07)$ is the quenching factor for \textalpha's~\cite{thedeapcollaborationRelativeMeasurementExtrapolation2024}, 
and $\eta_{\text{source}}=0.38(0.02)$ is the fraction of photons that leave the \textalpha\ source holder.

The expected light yield is calculated as $\text{LY} = Y \cdot \eta_{\rm det} \cdot \epsilon_{\rm PMT} \cdot \epsilon_{\rm PEN}$, where
$Y=$~\SI{40\pm8}{ph/\kilo\eV_{ER}} is the scintillation yield for electron recoil interactions in LAr~\cite{szydagis_2018_1314669,dokeEstimationAbsolutePhoton1990,dokeAbsoluteScintillationYields2002,segretoPropertiesLiquidArgon2021}, 
$\epsilon_{\rm PMT}$=17(1)\% is the quantum efficiency of the PMT~\cite{Babicz_2018}, and $\epsilon_{\rm PEN}$ is the PEN conversion efficiency. $\eta_{\rm det}$ is the fraction of photons that are not absorbed (i.e.~that eventually would reach the PMT if $\epsilon_{\rm PEN}$=1).
We determine $\eta_{\rm det}$ using Monte Carlo simulations. A model of the \textalpha\ source and its holder, as well as the cage layout was implemented in GEANT4. Literature values are used for the wavelength-dependent reflectivity of the WLSR foil~\cite{boulayDirectComparisonPEN2021} and for the Rayleigh scattering length of LAr~\cite{westerdaleDEAP3600LiquidArgon2024}. The absorption length for visible light is longer than \SI{60}{\meter}, as verified using the white LED. A VUV absorption length of approximately \SI{60}{\centi\meter} is most consistent with our data, including the measured LAr triplet lifetime.

For the \textalpha\ source located at the centre of the detector, we obtain $\eta_{\rm det}=1.3\%$. CR \textmuon's passing vertically through the detector have $\eta_{\rm det}=1.5\%$ in the center and $\eta_{\rm det}=1.8\%$ near the wall. 
As a cross-check, consistency with the  analytic model from ref.~\cite{segretoAnalyticTechniqueEstimation2012} was confirmed.

\section{Results} \label{sec:results}

\subsection{Time-evolution of the light yield} \label{sec:triplet_evolution}

The observed lifetime of the LAr triplet state, as well as the photon yield, are sensitive to impurities in the LAr~\cite{Acciarri:2009dj,jonesMeasurementAbsorptionLiquid2013a}. The photon yield is reduced when the energy of a LAr triplet excimer is passed to an impurity state, and when impurities absorb the VUV scintillation or the wavelength-shifted light. The effect of variations in impurities directly taking the excimer energy is accounted for by applying a triplet-lifetime-dependent correction factor to the PE yield for each day. The factor is defined as
$
\eta_i = ( \fpromptm + (1 - \fpromptm) \cdot \tau_0/ \tau_i),
$
where $\tau_0$ is the triplet lifetime of the first run, $\tau_i$ is the triplet lifetime of the $i^{\mathrm{th}}$ run, and we use \fprompt=0.78, which corresponds to the median of the first run's \fprompt~distribution. The drift in the triplet lifetime is not large enough to significantly affect the median \fprompt~value of the \textalpha\ population, thus no further correction is necessary.

\begin{figure}
    \centering
        \includegraphics[width=0.6\columnwidth]{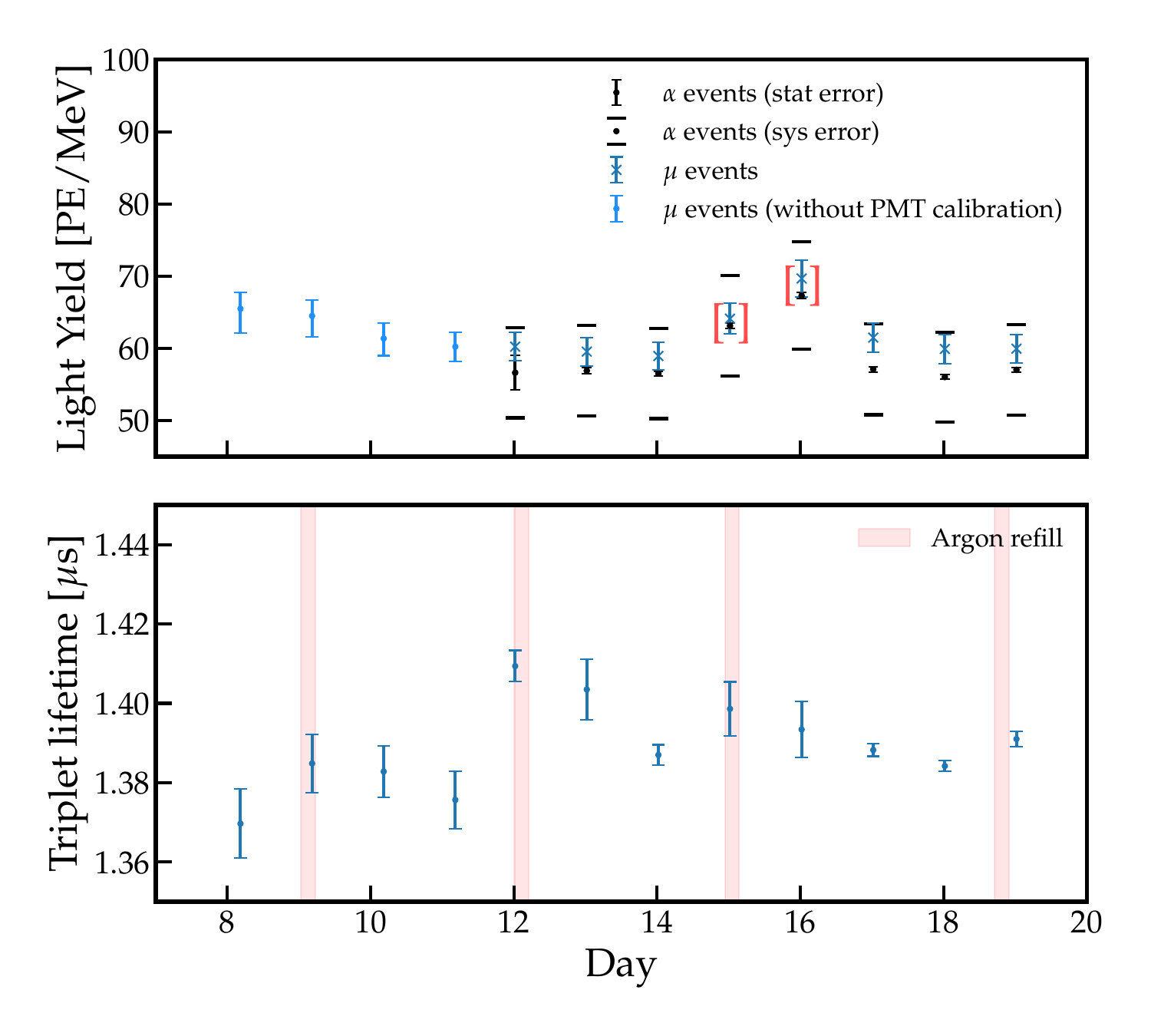}~\includegraphics[width=0.3\columnwidth]{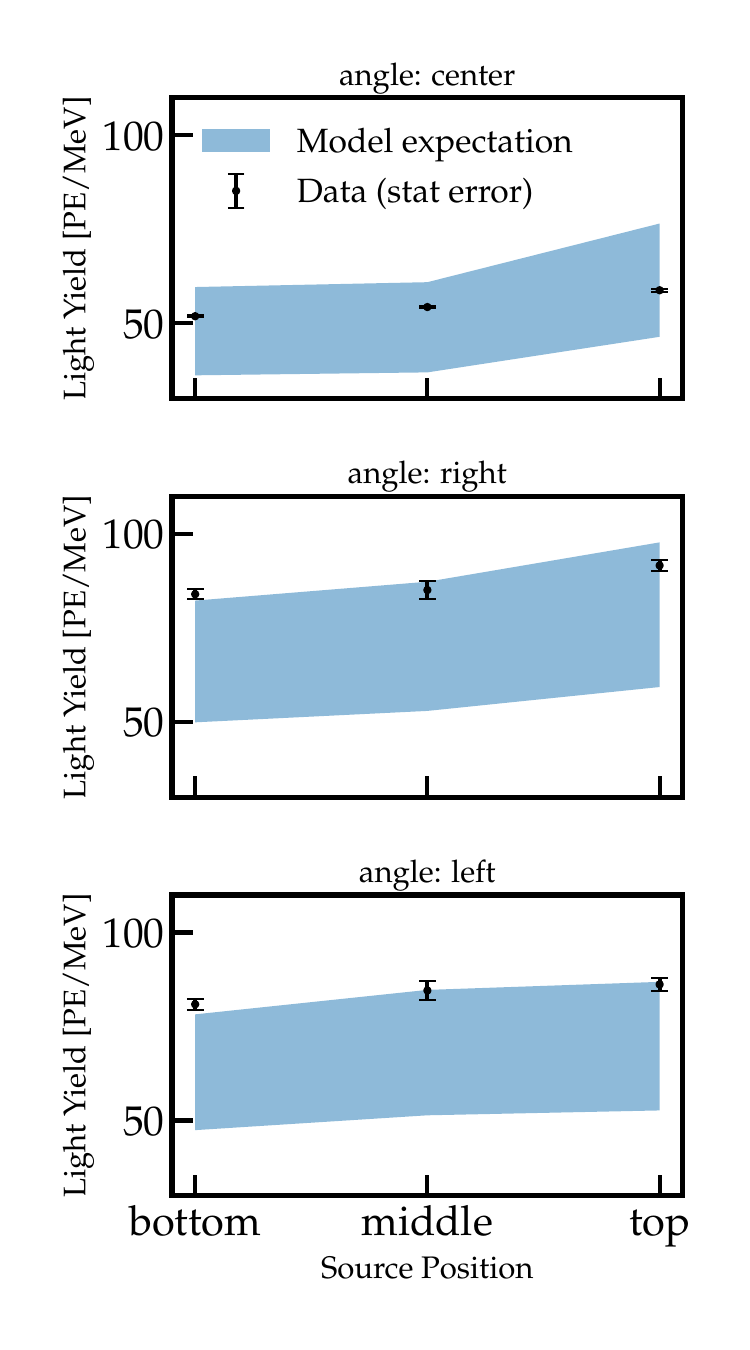}
    \caption{(left) Evolution of LY (after correcting for triplet lifetime variation) and the triplet lifetime. Data points in brackets were taken during periods of high noise rates. (right) Measured LY and the MC model expectation for different \textalpha\ source positions.\label{fig:qtotal-triplet}}
\end{figure}

The daily LYs are shown in figure~\ref{fig:qtotal-triplet}~(left) after correcting for triplet lifetime variation. Data taken with an oscilloscope during days 8 and 11 only allowed for the observation of the high-energy feature in the \textmuon\ spectrum, and PMT gain calibration was not possible in this time period. On days 15 and 16, a sudden unexplained high rate of correlated noise caused an apparent increase in the LY, and we exclude these two data points from analysis. 

\subsection{Variation of observed light yield with source position}
For the data shown in figure~\ref{fig:qtotal-triplet}~(left), the \textalpha\ source was placed at "top" (\SI{80}{\centi\metre}), centered horizontally. The LY measured at different locations inside the PEN cage is shown in figure~\ref{fig:qtotal-triplet}~(right). The expected LY at each source position is shown as shaded bands, where we used eq.~\ref{eq:ly} with $\epsilon_{\rm PEN}=0.45$. The width of the bands reflects systematic uncertainties.  

\section{Discussion and Conclusion} \label{sec:conclusion}
Our previous tabletop measurements have shown that PEN has a 47\% $\pm$ 6\% WLS efficiency relative to TPB~\cite{boulayDirectComparisonPEN2021}. Encouraged by these results, we set up \SI{4}{\square\metre} of combined PEN and specular reflector foils in a two-tonne LAr dewar to assess the stability over two weeks of detector operation. Given the \textalpha\ source activity and the rate and spectrum of cosmic \textmuon's, the VUV light exposure of the foil over this time frame is equivalent to 1\textendash3 years operation in a deep-underground rare-event search detector, in the case where VUV exposure is a main driver of light yield degradation.

The \textalpha\ and \textmuon\ LYs show excellent stability over eight days of measurement between days 12 and 19 and a linear fit estimates a slope consistent with zero ($ -0.0002 \pm 0.0012$~\% per day for \textalpha\ and $0.0015 \pm 0.0052 $~\% per day for \textmuon\ data). From the second dataset of \textmuon's between days 8 and 11, some decline in LY is visible. It could be caused by the degradation of the PEN response or impurity buildup due to a lack of active LAr circulation in the setup, particularly from impurities that absorb VUV light. If we assume that the decline is entirely due to PEN degradation, a linear fit estimates LY loss of $2.3 \pm 1.2 $~\% per day.

The LY from \SI{4.8}{\mega\eV} \textalpha's and \SI{300}{\mega\eV} cosmic \textmuon's shows agreement within the systematic error of eq.~\ref{eq:ly}. In contrast to the \textalpha\ source, the \textmuon's produce scintillation photons in the complete volume as they travel through the cage. Thus, the agreement in LY confirms the response of the PEN foils is uniform in the entire cage volume. The agreement in LY between the left and right positions of the \textalpha\ source further indicates a uniform response of the PEN foils.

There is a large uncertainty on the expected LY due to the uncertainties in the LAr scintillation parameters and the various efficiencies. We observe a difference in LY between the center and the edges of the detector that cannot fully be reproduced in simulation. Matching the simulation to the center data points leads to a PEN efficiency of approximately 45\%. If we match to the edge data instead, the efficiency would be 55\%. The PEN efficiency is highly degenerate with the VUV absorption length, for which we took a rather low value in order to best match the data points at all positions. The discrepancy between center and edge likely has to do with the simulation not reproducing the angular distribution of the wavelength-shifted photons in a realistic manner. Either way, these PEN efficiencies are consistent with past measurements~\cite{efremenkoUsePolyEthylene2020,abrahamWavelengthShiftingPerformancePolyethylene2021}.

These results show PEN is a viable option for scintillation light detection in the future large scale LAr detectors and the ease of instrumentation with PEN would be a serious advantage in simplifying the photon detection system of the next generation of LAr experiments.

\acknowledgments
This work was funded by the EU’s Horizon 2020 research and innovation programme under grant agreement No 962480 (DarkWave project) and grant agreement No 101004761, the International Research Agenda Programme AstroCeNT (MAB/2018/7) funded by the Foundation for Polish Science from the European Regional Development Fund (ERDF), the Swiss National Science Foundation (Grants No.20FL20-216572 and No.200020-219290), the UZH Postdoc Grant No.K-72312-14-01, and the Royal Society UK. K.~Thieme was supported by the U.~S.~National Science Foundation~(NSF) (Grants No.~PHY-1812482 and PHY-2310046). A.~Leonhardt was supported by the Deutsche Forschungsgemeinschaft (DFG, German Research Foundation) through the Sonderforschungsbereich (Collaborative Research Center) SFB1258 ‘Neutrinos and Dark Matter in Astro- and Particle Physics’.  The prototype construction was carried out at the CEZAMAT (Warsaw University of Technology) cleanroom infrastructures financed by the ERDF; we thank the CEZAMAT staff for support. Support from the CERN Neutrino Platform is gratefully acknowledged.
\bibliography{cernwls_bibliography}
\bibliographystyle{tp_unsrt_doi}

\end{document}